\documentclass[
 amsmath,amssymb,
 aps, physrev,
 reprint,%
]{revtex4-2}

\usepackage{graphicx}
\usepackage{dcolumn}
\usepackage{bm}
\usepackage{hyperref}
\usepackage{color}
\usepackage[caption=false]{subfig}
\usepackage{breakcites}
\begin{document}

\preprint{APS/123-QED}

\title{Using Statistical Mechanics to Improve Real-World Bayesian Inference: A New Method Combining Tempered Posteriors and Wang-Landau Sampling}


\author{Alfred C.K. Farris}
 \email{Contact author: afarris@lanl.gov}
  \affiliation{Los Alamos National Laboratory, Los Alamos, New Mexico 87545, USA
}

\begin{abstract}
\noindent We present a simple method to obtain optimal posterior distributions and improve the quality of Bayesian inference with reduced human and computational effort. Bayes' Theorem is reformulated in the language of statistical mechanics, wherein an improved posterior -- referred to as a tempered posterior -- is defined analogously to a canonical probability distribution at temperature $\tau$. Wang-Landau sampling is used to obtain the density of states of the posterior probability, and signals analogous to those of phase transitions are extracted from a single simulation. In addition, the transition temperature is easily identified, providing the tempered posterior with optimal predictive performance. We demonstrate the efficacy of the method on a real-world problem in materials science (equation of state modeling) with messy data, a high-dimensional and correlated input parameter space, and ``frustration'' among model outputs.

\end{abstract} 

\maketitle

Across disciplines, Bayesian inference is used to perform model uncertainty quantification. This process utilizes existing data to obtain probability distributions of model input parameters such that model outputs accurately predict future observations \cite{gelman2013}. Bayes' Theorem provides a recipe to obtain the posterior distribution $\mathcal{P}(\theta|D)$, which specifies the probability of model parameters $\theta$ conditional on the data $D$:
\begin{equation}\label{eq:posterior}
    \mathcal{P}(\theta|D)= \frac{L(D|\theta)\pi(\theta)}{\int L(D|\theta)\pi(\theta)d\theta}.
\end{equation}

\noindent The likelihood $L(D|\theta)$ parametrizes the relationship between the data and physical model, the prior $\pi(\theta)$ represents our initial beliefs about model parameters, and the denominator is the normalizing constant called the model evidence. 



In real-world applications there is no ``correct'' prior or likelihood, due to the complex relationship between imperfect data and imperfect models. This leads to numerous practical difficulties when implementing the Bayesian framework, centered around the time-intensive and potentially error-prone process of model refinement and calibration. Human-time is needed to design and evaluate iterative improvements to priors and/or likelihoods \cite{vandeSchoot2021}, and since the model evidence is analytically intractable, extensive computer-time is spent performing Monte Carlo sampling to obtain each iterative estimate of the posterior. Moreover, the most commonly used Monte Carlo acceptance criterion (Metropolis-Hastings \cite{Metropolis1953,hastings,vandeSchoot2021,zuckerman2024}) suffers from systematic errors due to well-known sampling limitations \cite{Landau_Binder_2014}. 

Tempered posteriors $\mathcal{P}_{\tau}(\theta|D)$ have emerged as a potential alternative to standard model refinement. In this case, rather than redefining the likelihood and/or prior, the goal is to determine the fictitious temperature $\tau$ which is introduced to either narrow ($\tau < 1$) or widen ($\tau > 1$) the untempered ($\tau=1$) distribution:
\begin{equation}\label{eq:coldposterior}
    \mathcal{P}_{\tau}(\theta|D)= \frac{[L(D|\theta)\pi(\theta)]^{1/\tau}}{\int[L(D|\theta)\pi(\theta)]^{1/\tau}d\theta}.
\end{equation}
A tempered posterior is a proper Bayesian posterior with an alternative prior and likelihood \cite{zhang2024}, and thus this tempering procedure offers a promising surrogate for Bayesian model refinement.
Cold posteriors ($\tau < 1$) have improved predictive performance in interatomic potential model calibration in physics \cite{frederiksen2004}, Bayesian linear regression \cite{zhang2024},  and Bayesian neural networks \cite{wenzel2000}. Tempering alleviates underfitting in cases where the original likelihood and/or prior were misspecified \cite{zhang2024}. 

An attractive feature of this approach is its link to statistical mechanics. The significance of tempering the posterior and the connection between $\tau$ and a physical temperature has been frequently explored in recent years \cite{skilling2006,habeck12, Murayama2024}. For analytic models, it has been shown that $\tau$ at which optimal posterior predictive performance occurs is marked by signals akin to phase transitions in physical systems \cite{lamont2019,Murayama2024}. However, obtaining these analogous signals remains extremely challenging in all but ideal cases \cite{plummer2026specheat,plummer2026thermo}.

In this Letter, we present a method which allows for the use of tempered posteriors to improve Bayesian model predictive performance without the need for model refinement or resampling. We use Wang-Landau sampling \cite{Wang2001,Wang2001_2} to systematically explore and characterize the complex posterior landscape, bypassing issues which plague  traditional methods. Treating this landscape as analogous to a complex landscape in a physical system, we then show how standard relations from statistical mechanics can be used to obtain tempered posterior distributions for \textit{all} values of $\tau$ from a single simulation. This allows us to easily extract the optimal tempered posterior with improved predictive performance by identifying the aforementioned ``phase transitions'' in the Bayesian system, where relevant signals also correspond to the maximum of the Fisher information. Our approach is firmly grounded in statistical physics and information theory, providing a quantitative alternative to the standard ad-hoc and time-consuming procedure of model refinement. We demonstrate the utility of our method by performing uncertainty quantification of an analytic equation of state (EOS), a key ingredient in materials modeling. This application exhibits difficulties commonly found in real-world inference -- messy data from which to fit approximate models, a high-dimensional and correlated input parameter space, and ``frustration'' among model outputs.

At the heart of our method is the change of variables:
\begin{equation}
    \mathcal{P}_\tau(\theta|D)= P_\tau(E=-\ln\left[L(D|\theta)\pi(\theta)\right]),\label{eq:ptpe}
\end{equation}
\noindent where
\begin{equation}
    P_{\tau}(E)=\frac{g(E)e^{-E/\tau}}{\int g(E)e^{-E/\tau}dE},\label{eq:pe}
\end{equation}
\noindent with
\begin{equation}
    g(E)=\int\delta\{E+\ln\left[L(D|\theta)\pi(\theta)\right]\} d\theta.\label{eq:ge}
\end{equation}
\noindent Remarkably, Equation \eqref{eq:pe} is analogous to a canonical distribution at temperature $\tau$ of a physical system characterized by a density of states $g(E)$. Shifting our focus away from directly sampling the posterior distribution at a particular temperature $\tau$ as is done traditionally, we note that by obtaining $g(E)$ from Wang-Landau sampling \cite{Wang2001,Wang2001_2}, the tempered posterior distribution is immediately available at \textit{all} values of $\tau$ from a single simulation. 

In further analogy to a physical system where the degrees of freedom and microstate configuration determine the energy of a macrostate, in the Bayesian framework the model parameters $\theta$ of a ``microstate'' determine the ``macrostate'' $E(\theta)$. In both cases, the density of states quantifies the number of microstates which correspond to a given macrostate. Our definition of $g(E)$ differs slightly from similar approaches \cite{skilling2006,habeck12} and allows for a more direct parallel with the density of states in statistical physics. Moreover, our definition lends itself to the use of Wang-Landau sampling, whose ease of implementation has led to broad adoption in the computational statistical physics community. Note that the negative sign in Equation \eqref{eq:ptpe} is present because while lower values of the energy are favorable in a physical system, higher values of the probability $L(D|\theta)\pi(\theta)$ are favorable in Bayesian inference.

\begin{figure}[t!]
\includegraphics[width=0.48\textwidth]{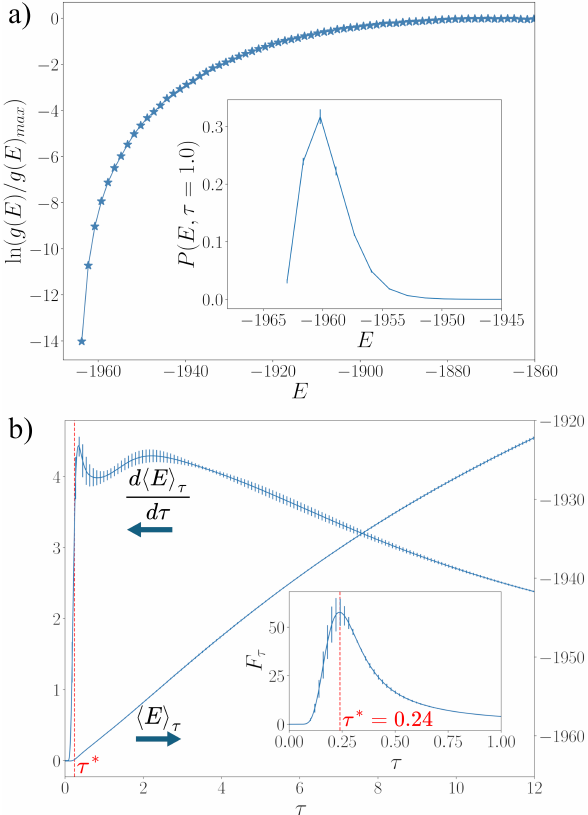}
\caption{\label{fig:thermo} (a) Density of states $g(E)$ with $E(\theta)=-\ln\left[L(D|\theta)\pi(\theta)\right]$ obtained from Wang-Landau sampling. Even with only 8 parameters, $g(E)$ spans many orders of magnitude. The inset shows $P(E,\tau=1.0$) obtained from Equation \eqref{eq:pe}. $P(E)$ can be used to determine which values of $E$ (and thus which parameter sets $\theta$) have greater than $1\%$ probability at a given $\tau$. (b) $\left<E\right>_{\tau}$ obtained from Equation \eqref{eq:eavg} and its first derivative $d\left<E\right>_{\tau}/d\tau$. As $\tau$ increases, portions of the posterior landscape away from the global minimum become accessible, thus increasing $\left< E \right>_{\tau}$. The first derivative is the analogue of the heat capacity in physical systems, and signals indicate the equivalent of phase transitions. The inset shows the Fisher information $F_{\tau}$, obtained from Equation \eqref{eq:Fish}. The location of the maximum Fisher information signifies the critical temperature $\tau^*=0.24$, corresponding to the tempered posterior with optimal predictive capability. Note that this critical temperature is slightly below the low temperature transition signal in $d\left<E\right>_{\tau}/d\tau$. Standard errors were obtained from 4 independent Wang-Landau simulations; error bars smaller than the size of the points are not shown.}
\end{figure} 

In statistical mechanics, moments of the canonical probability distribution carry meaning, and we will use our understanding of this meaning to analyze the tempered Bayesian posterior. From $g(E)$, it is straightforward to compute moments (of order $n$) at any $\tau$:
\begin{equation}
    \left<E^n\right>_{\tau}=\frac{\int E^ng(E)e^{-E/\tau}dE}{\int g(E)e^{-E/\tau}dE}. \label{eq:eavg}
\end{equation}

\noindent The first and second moments are particularly important, as they are related to the heat capacity in physics ($d\left< E\right>_{\tau}/d{\tau}$) and the Fisher information $F_{\tau}$ in information theory \cite{headley2026quantumfisherinformationentropy}:
\begin{equation}
    F_{\tau}=\frac{1}{\tau^2}\frac{d\left< E\right>_{\tau}}{d\tau}=\frac{\left< E^2\right>_{\tau} - \left< E\right>_{\tau}^2}{\tau^4}.\label{eq:Fish}
\end{equation}
\noindent  The Fisher information is often used as a quality metric for parameter estimation in model fitting -- the higher $F_{\tau}$, the more information a fitting parameter contains about the data \cite{Frieden_2004}. Signals in the heat capacity and Fisher information are also used to identify phase transitions in computer simulations of finite physical systems \cite{Landau_Binder_2014,janke2004}. We will show below that these two quantities are essential for improving the quality of our Bayesian inference.

\begin{figure}
\includegraphics[width=0.48\textwidth]{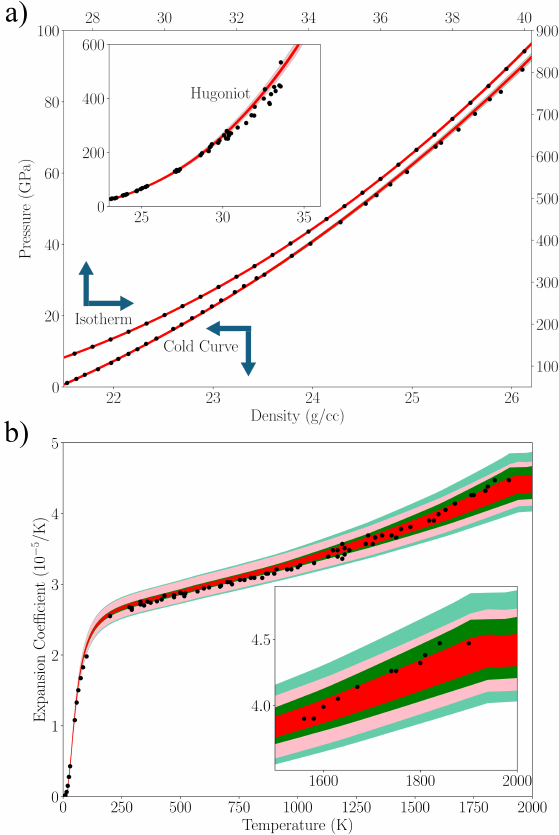}
\caption{\label{fig:physicsEOS} Model output from posterior distributions compared to experimental data (black circles). The red and green curves correspond to $\tau^*=0.24$ and $\tau=1.0$, respectively, while the dark and light shades correspond to the EOS curve and the EOS curve plus $\sigma$, respectively. 3000 samples are shows for each posterior. The model-output expansion coefficient (b) shows drastically different behavior between posteriors. The untempered posterior clearly underfits the data, while the posterior obtained from the critical temperature $\tau^*$ provides a better fit and thus better model predictive performance.}
\end{figure}

Our application in this work is uncertainty quantification of the input parameters for an analytic EOS model for the FCC phase of platinum. An EOS is a model for a material's thermodynamic properties, and plays an important role in the materials modeling. For the purpose of this work, the problem is best viewed as a high-dimensional curve fit of approximate physical models with correlated input parameters to messy experimental data. Our fit is constrained by $j=4$ sets of experimental data with $i_j$ data points each, shown as black circles in Figure \ref{fig:physicsEOS}. Experimental data are obtained from \cite{PhysRevB.70.094112,White1972,Hahn1972,Austin1932,CINDAS_TPMD,Arblaster1997,Marsh1980,Holmes1989,Cochrane2022}, and we assume for simplicity that they are Gaussian distributed about the model. We aim to estimate the posterior probability of four EOS model input parameters $\{\Lambda\}$ and the Gaussian variance for each data set $\{\sigma'\}$, which we assume grows in proportion to the y-value of the data as in \cite{lindquist2022}: $\sigma'_j = \sqrt{2}y_{ji_j}\sigma_j$. This defines the likelihood:
\begin{equation}
    L(\theta)=\prod_{j}\prod_{i_j}\frac{1}{\sigma^*_{j}\sqrt{2\pi}}\mathrm{exp}\left( \frac{-(y_{ji_j}-y_{ji_j}^{model}(\{\Lambda\}))^2}{2\sigma^{*2}_{j}}\right).
\end{equation}
We adopt Jeffreys prior \cite{jeffreys} for the variance terms so that $P(\sigma) \propto 1/\sigma$. Note that the proportionality and normalization constants don't matter since we only care about relative probabilities. Our EOS is generated using the OpenSESAME production code at Los Alamos National Laboratory; see \cite{mchardy2018} and references therein for more information about EOS theory and the code. The details of this particular EOS are planned for a future publication. 

\begin{figure}
\includegraphics[width=0.48\textwidth]{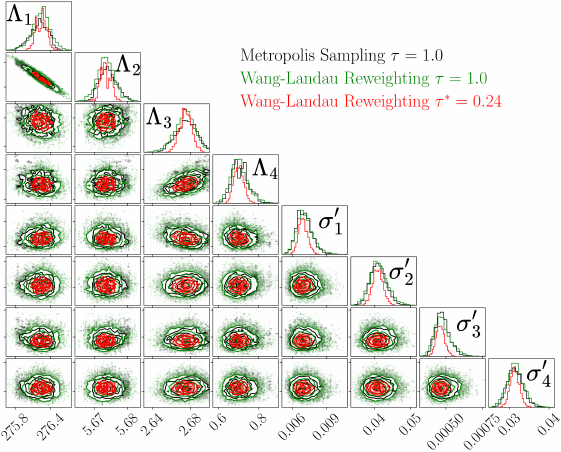}
\caption{\label{fig:corners} (On-diagonal) Histograms of the model parameters from Wang-Landau sampling for both the critical posterior $\tau^*=0.24$ (red) and the untempered posterior $\tau=1.0$ (green). For comparison, we show results from a simple Metropolis sampling (black) of the untempered posterior; the Metropolis results overlap almost exactly with our method of reweighting $g(E)$ to $\tau=1.0$, providing a quick and easy sanity-check. Since $\tau^* < 1.0$, the critical posterior narrows the model parameter ranges. Each histogram contains on the order of $10^4$ samples. (Off-diagonal) Pairwise marginal distributions which highlight correlations between model parameters. These also provide another perspective on the relative width of the posterior at both values of $\tau$.}
\end{figure}

Our Monte Carlo sampling proceeds sequentially through the input and variance parameters and consists of both global and local moves. Each Monte Carlo trial move has a $10\%$ chance of being global (such that the walker can move anywhere within the predefined parameter range) and a $90\%$ of being local (defined as the current parameter value plus or minus a uniform random fraction of $5\%$ of the parameter range). For the single-walker Wang-Landau sampling, we adopt a flatness criterion of $60\%$, an initial modification factor of $f_0=e^1$, and a minimum modification factor of $f_{min}=\mathrm{exp}(10^{-8})$, with $f_{k+1} = \sqrt{f_k}$. Standard error bars were obtained from four independent Wang-Landau simulations.

The density of states $g(E)$ obtained from Wang-Landau sampling for this system is shown in Fig. \ref{fig:thermo}a; even with only 8 parameters, $g(E)$ spans many orders of magnitude -- further increasing the dimensionality will make this an even harder problem. Figure \ref{fig:thermo}b shows $\left<E\right>_{\tau}$ and its first derivative with respect to $\tau$. The highest probability parameter set occurs at $\tau=0.0$  and, as expected, $\left<E\right>_{\tau}$ increases with $\tau$ as regions of the posterior landscape with worse fits become accessible. This is analogous to sampling a physical system around its global minimum (ground state).

Signals in $d\left< E\right>_{\tau}/d{\tau}$ provide insight into the curvature of the posterior landscape and indicate the equivalent of phase transitions in physical systems \cite{lamont2019,Murayama2024}. We suspect that the higher temperature peak corresponds to a loss of model sloppiness, as was observed in the equivalent ``learning capacity'' in \cite{lamont2019}, but the high-temperature signal is not the primary focus of this work. Given the striking similarity between this curve and the heat capacity in simple models for protein folding \cite{farris2019}, we infer that the low temperature peak corresponds to an abrupt decrease in the entropy -- relative to all possible parameter sets $\theta$, there are only a few for which the model fits the data. This is analogous to how a protein has many possible configurations, but only a few which are similar enough to the ground state for the protein to function as intended. Providing further justification for this insight is the peak corresponding to the maximum Fisher information $F_{\tau}$, shown in the inset of \ref{fig:thermo}b. The temperature at which the maximum of $F_{\tau}$ occurs corresponds to the temperature at which the model captures the most possible information relative to the data, or, in other words, the model can best reproduce the data and make optimal predictions. Thus, we define this temperature as the critical temperature $\tau^*=0.24$. Note that $\tau^*$ is slightly below the low temperature transition signal in $d\left< E\right>_{\tau}/d{\tau}$ -- this is comforting, since sampling exactly at the temperature of the peak in $d\left< E\right>_{\tau}/d{\tau}$ would provide an undesirable distribution with states both above and below the transition.

We now wish to compare model parameters between the untempered posterior $\tau=1.0$ and the critical posterior $\tau^*=0.24$. To obtain parameter distributions, we first compute $P(E)$ using Equation \eqref{eq:pe}, shown for $P(E,\tau=1.0$) in the inset of Figure \ref{fig:thermo}a. We save model parameters for every value of $E + dE$ we encounter during the Wang-Landau sampling ($dE$ is the small but finite bin width). Thus, we can identify all sets of parameters $\theta$ for which $P(E(\theta)) > 0.01$ (our chosen threshold). Figure \ref{fig:corners} shows histograms of the model parameters for both $\tau^*=0.24$ and $\tau=1.0$ (on the order of $10^4$ samples for each case), as well as pairwise marginal distributions which highlight correlations between model parameters. For comparison, we also show results from a simple Metropolis sampling of the untempered posterior using the same Monte Carlo trial moves specified previously. The Metropolis results overlap with our method of reweighting $g(E)$ to $\tau=1.0$, providing a quick and easy sanity-check. Since $\tau^* < 1.0$, narrowed parameter distributions are found for the critical posterior.

The effect on model predictive capability is apparent when we plot outputs from the posterior for both $\tau^*=0.24$ and $\tau=1.0$, shown in Figure \ref{fig:physicsEOS} ($3000$ samples each). The red and green curves correspond to $\tau^*=0.24$ and $\tau=1.0$, respectively, while the dark and light shades correspond to the EOS curve and the EOS curve plus $\sigma$, respectively. While most curves overlap on this scale,  the expansion coefficient (Figure \ref{fig:physicsEOS}b) shows distinctly different behavior between posteriors, and it is clear that the untempered posterior underfits the data. Rather than performing an expensive model refinement and resampling as is done traditionally, identification of the critical posterior and simple postprocessing provide us with a posterior distribution with \textit{much} better predictive capability, essentially for free. The reason we are still capturing more than the $68\%$ of the data appropriate to the one-sigma measure can be attributed to the correlations between model parameters and outputs, the assumption that $\sigma$ should grow in proportion to the $y$-value of the data, and the fact that we assume the data are Gaussian distributed about the model, which is clearly not the case in the Hugoniot (inset of Figure \ref{fig:physicsEOS}a). However, these kinds of limitations are unavoidable in real-world, non-idealized systems -- for example, attempting to infer $\sigma$ for each data point is simply intractable given the huge increase in dimensionality the sampling would incur. The benefit here is that given a likelihood and prior we believe to be appropriate for the problem, we can make large improvements to model predictive capability without the traditional expensive loop of model refinement and resampling.

We believe the approach outlined in this Letter has the potential to produce significant improvements to model predictive performance for real-world problems. From a fundamental perspective, we also believe our approach improves on the current, slightly ad-hoc process by which Bayesian inference is done -- the critical temperature associated with phase transition signals and the maximum Fisher information provides a quantitative measure with which to assess the quality of Bayesian inference. This study also opens many avenues for exploration. For low-hanging fruit, small but important gains in predictive performance may be obtained from simple histogram reweighting \cite{Landau_Binder_2014} of the untempered posterior readily available using existing codes and approaches. Another interesting possibility is to explore these transition signals explicitly in terms of the entropy using the microcanonical entropy and its inflection point analysis \cite{Qi2018}. This could provide additional quantitative measures by which to judge the quality of Bayesian inference. Since it is straightforward to compute the model evidence from $g(E)$, Wang-Landau sampling (or a recently introduced adaptive version for sampling posterior landscapes \cite{bornn2013}) could provide a simple alternative to Nested sampling \cite{skilling2006} or thermodynamic integration \cite{gelman1998} for performing Bayesian model selection \cite{llorente2023}.  \\
\indent \textit{Acknowledgments} -- We thank Effrosyni Seitaridou, J\'{e}r\^{o}me Daligault, Scott Crockett, and Sven Rudin for the many helpful discussions. This work was supported by the U.S. Department of Energy National Nuclear Security Administration under Contract No. 89233218CNA000001. Approved for unlimited release LA-UR-26-22974.

\textit{Data availability} -- Data are available upon reasonable request.




\bibliography{WL_EOS}

\end{document}